\documentclass{aa}

\makeatletter
\let\linenumbers\relax
\makeatother

\usepackage[varg]{txfonts}
\usepackage{graphicx}
\usepackage{gensymb}
\usepackage{color}
\usepackage[colorlinks=true,citecolor=blue,linkcolor=blue]{hyperref}
\usepackage{xspace}
\usepackage{CJKutf8}
\usepackage{comment}

\newcommand{\chandra}{\textsl{Chandra}\xspace}
\newcommand{\nustar}{{NuSTAR}\xspace}
\newcommand{\xmmnewton}{{XMM-\textsl{Newton}}\xspace}

\newcommand{\xrism}{{XRISM}\xspace}
\newcommand{\nicer}{{NICER}\xspace}
\newcommand{\newathena}{\textsl{NewAthena}\xspace}

\begin{document}

\title{First look at Vela X-1 with \xrism}
\subtitle{A simultaneous campaign with \xmmnewton and \nustar}

\author{
C.~M.~Diez\inst{\ref{affil:ESAC}}\and
S.~Dupourqué\inst{\ref{affil:IRAP}}\and
M.~Zhou \begin{CJK*}{UTF8}{gbsn}(周孟磊)\end{CJK*}\inst{\ref{affil:IAAT}}\and
E.~Quintin\inst{\ref{affil:ESAC}}\and
G.~A.~Matzeu\inst{\ref{affil:Quasar}}\and
F.~Fürst\inst{\ref{affil:ESAC}}\and
P.~Kretschmar\inst{\ref{affil:ESAC}}\and
R.~Amato\inst{\ref{affil:OAR}}\and
C.~Malacaria\inst{\ref{affil:OAR}}
}

\institute{European Space Agency (ESA), European Space Astronomy Centre (ESAC), Camino Bajo del Castillo s/n, 28692 Villanueva de la Cañada, Madrid, Spain \label{affil:ESAC} \\
email: \texttt{camille.m.diez@gmail.com}
\and
IRAP, Université de Toulouse, CNRS, CNES, UT3-PS, Toulouse, France \label{affil:IRAP}
\and
Institut für Astronomie und Astrophysik, Universität Tübingen, Sand 1, 72076 Tübingen, Germany \label{affil:IAAT}
\and
Quasar Science Resources SL for ESA, European Space Astronomy Centre (ESAC), Science Operations Department, 28692 Villanueva de la Cañada, Madrid, Spain \label{affil:Quasar}
\and
INAF-Osservatorio Astronomico di Roma, Via Frascati 33, I-00078 Monte Porzio Catone, (RM), Italy \label{affil:OAR}
}

\date{Received 3 July 2025 / Accepted 3 September 2025}

\abstract{High-mass X-ray binaries (HMXBs) can serve as useful laboratories for exploring the behaviour of accreted matter onto compact objects and for probing the complex wind environments of massive stars. These investigations are essential for understanding stellar life cycles and the dynamics of the Milky Way, and they are prominent topics in the science cases for \xrism and \newathena.}{We report, for the first time, on a \xrism observation of the HMXB Vela X-1, conducted during the first cycle of the \xrism general observer programme and complemented by simultaneous \xmmnewton and \nustar coverage. This campaign targeted a critical orbital phase -- when the neutron star is in inferior conjunction -- during which significant changes in absorption are expected.}{We performed absorption-resolved spectral analyses during two time intervals of interest: the soft and hard hardness ratio (HR) intervals, as it is strongly correlated with absorption variability.}{We observed a sudden transition in the HR from a soft to a hard state coinciding with an increase in the absorption column density. This is likely attributable to the onset of the accretion structure crossing our line of sight. With \xrism/Resolve, we also investigated the Fe K region, and we report for the first time the presence of a Fe K$\alpha$ doublet in the spectrum of Vela X-1 together with the presence of already known Fe K$\beta$ and Ni K$\alpha$ lines that are produced in cold clumps embedded in the hot ionised wind. The measured line velocities of the order of $10^2 \ \mathrm{km\,s^{-1}}$ are consistent with production sites in the vicinity of the neutron star.}{This precursor study with Vela X-1 shows the potential of \xrism to study in unprecedented detail the spectral evolution of wind-accreting X-ray binaries.} 


\maketitle

\section{Introduction}
\label{section:intro}

\begin{figure}[htpb]
    \centering
    \centerline{\includegraphics[trim=0cm 0cm 0cm 0cm, clip=true, width=0.9\linewidth]{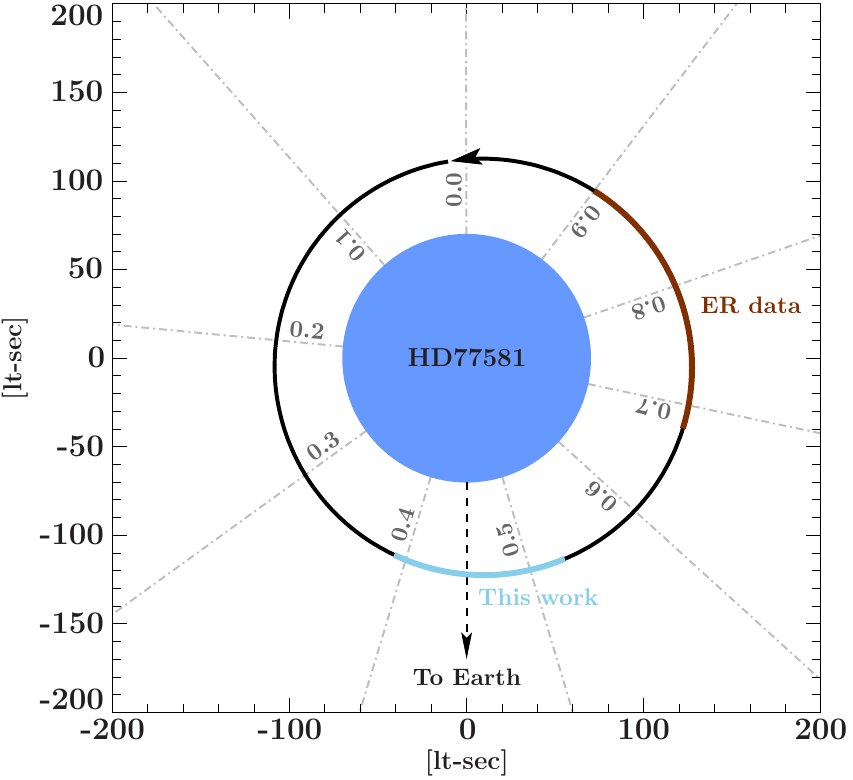}}
    \caption{Sketch of the Vela X-1 system showing the orbital phases covered by the neutron star during our \xrism observation (light blue) and during the ER observation (dark brown). In this image, the observer is facing the system at the bottom of the picture, and $\phi_{\rm{orb}} = 0$ is defined at the eclipse.}
    \label{fig:orbit}
\end{figure}

Vela X-1 is an archetypal eclipsing high‑mass X‑ray binary (HMXB), and as it is located at only $1.99^{+0.13}_{-0.11}$ kpc away \citep{Kretschmar_2021}, it is one of the brightest persistent point sources in the X-ray sky despite its moderate average intrinsic luminosity of $\sim$5$\times10^{36}\,\mathrm{erg\,s^{-1}}$ \citep{Fuerst_2010a}. This system has been extensively reviewed by \citet{Kretschmar_2021}, who provide a comprehensive overview of its parameters. Vela X-1 comprises a B0.5 Ib supergiant star, HD 77581 \citep{Hiltner_1972}, and an accreting neutron star in a close quasi-circular orbit with a period of $\sim$8.964 days \citep{Kreykenbohm_2008, Falanga_2015}. A dense stellar wind with $\dot M\sim10^{-6}\,M_{\odot}\,\mathrm{yr}^{-1}$ \citep[see e.g. ][]{Watanabe_2006} feeds the neutron star, producing pulsed X‑ray emission with a variable period of $\sim$283 s. Edge‐on systems such as Vela X‑1 \citep[$i>73^\circ$,][]{vanKerkwijk_1995} enable orbital phase‐resolved spectroscopy to probe different structures and regions in the stellar wind that are modified by interaction with the neutron star. Variations in the hydrogen absorption column density, $N_{\mathrm{H}}$, on our line of sight -- a direct tracer of the stellar wind -- show a post‐eclipse decline until $\phi_{\mathrm{orb}} \approx 0.2$--$0.3$ \citep{Martinez_2014}. This decline in absorption is followed by a rise in the 0.4--0.6 range, and it was first detected by \cite{Ohashi_1984} and more recently monitored with high time resolution by \citet{Diez_2023}. At this range, the $N_{\mathrm{H}}$ seems to stabilise at high values until the eclipse \citep{Abalo_2024}. Local extrema attributed to the presence of clumps often modify this pattern and create rapid flux and spectral variability in Vela X‑1 \citep[][]{Grinberg_2017,Diez_2023}.
Hydrodynamic simulations \citep{Manousakis_2011} together with prior observations \citep[][]{Grinberg_2017} have shown that an accretion wake trails the neutron star, and this is possibly caused by its supersonic motion around the companion star. In addition, the X‑ray photoionisation from the compact object produces a large‐scale filamentary wake \citep{Blondin_1990a} known as the photoionisation wake. When these wakes intersect our line of sight, they absorb soft photons and reveal ionised emission lines in the Ne, S, Mg, and Si regions as well as Fe K$\alpha$, K$\beta$, and Ni K$\alpha$ fluorescent lines \citep{Sako_1999, Goldstein_2004, Martinez_2014, Amato_2021}. A public early release (ER) of a 64\,ks \xrism observation of Vela X-1\footnote{\url{https://heasarc.gsfc.nasa.gov/docs/xrism/results/erdata/index.html}. The ER data can be used to prepare proposals for observing calls, but reliable scientific results cannot be derived from these data products until the full data and calibration are available.} at $\phi_{\mathrm{orb}} \approx 0.68$--$0.90$ (see Fig.~\ref{fig:orbit}), not yet published in a journal, showcases fluorescent lines in the 6–8\,keV band, including the Fe K$\alpha_1$/K$\alpha_2$ doublet and Fe K$\beta$ line, and the ionised Fe XXV triplet.

In our targeted \xrism observation, we aimed to cover the inferior conjunction of the neutron star (see Fig.~\ref{fig:orbit}). At these phases, the passage of the accretion and ionisation wakes along the line of sight was observed in an earlier \xmmnewton and \nustar campaign \citep{Diez_2023}. Monitoring this orbital phase enables measurement of the change in absorption column density at low energies as the wakes cross our line of sight. To constrain the high-energy continuum and compensate for the energy range not accessible to \xrism/Resolve due to the closed gate valve (see Sect.~\ref{appendix:data_reduc} in the Appendix), we requested simultaneous observations with \nustar and \xmmnewton.

\section{Observations and data reduction}
\label{section:obs_data_reduc}

This coordinated observation of Vela X-1 with \xrism\ \citep{Tashiro_2025}, \nustar\ \citep{Harrison_2013}, and \xmmnewton\ \citep{Jansen_2001} was conducted on 19 November 2024. We used the European Photon Imaging Camera pn-CCD (EPIC-pn; \citealp{Strueder_2001}) operated in timing mode on \xmmnewton. For \nustar, data were obtained with the Focal Plane Modules A and B (FPMA and FPMB). Finally, regarding \xrism, we employed the high-resolution X-ray imaging spectrometer Resolve and the soft X-ray imager Xtend \citep{Tashiro_2025} configured in the 1/8 window mode with 0.1\,s burst mode. Further details of the observations are provided in Table~\ref{tab:obs_details} in the Appendix, and a sketch of the Vela X-1 system showing the orbital phases we covered is shown in Fig.~\ref{fig:orbit}. We present in Sect.~\ref{appendix:data_reduc} in the Appendix a short description of the specific data reduction that has been applied for \xmmnewton/EPIC-pn in timing mode and technical details encountered during the \xrism/Resolve data acquisition.

The data analysis was performed using \texttt{jaxspec} v0.3.0 \citep{Dupourque_2024}. This X-ray spectral analysis Python library leverages the use of differentiable approaches for Bayesian inference, such as variational inference \citep[see e.g.][]{Blei_2017}, thus enabling thawing of all the parameters of the spectral model, and it provides robust constraints within short timescales. The models were fitted with variational inference with a multivariate Gaussian guide for the posterior parameters and adjusted using the Adam optimiser \citep{kingma2017adammethodstochasticoptimization} with a learning rate of $10^{-3}$. We sought to minimise a Poisson likelihood for the observed spectrum, including a model-free background to account for its intrinsic uncertainty. We also used the Interactive Spectral Interpretation System (\texttt{ISIS}) v1.6.2-53 \citep{Houck&Denicola_2000} for comparison and sanity checks with \texttt{jaxspec} and for plots. Both \texttt{ISIS} and \texttt{jaxspec} provide access to \texttt{XSPEC} \citep{Arnaud_1996a} models, which are referenced later in the text.

\section{Results}
\label{section:results_discuss}
\subsection{Timing analysis}
\label{subsection:timing}

\begin{figure}[!htpb]
    \centering
    \centerline{\includegraphics[trim=0cm 0cm 0cm 0.5cm, clip=true, width=1.0\linewidth]{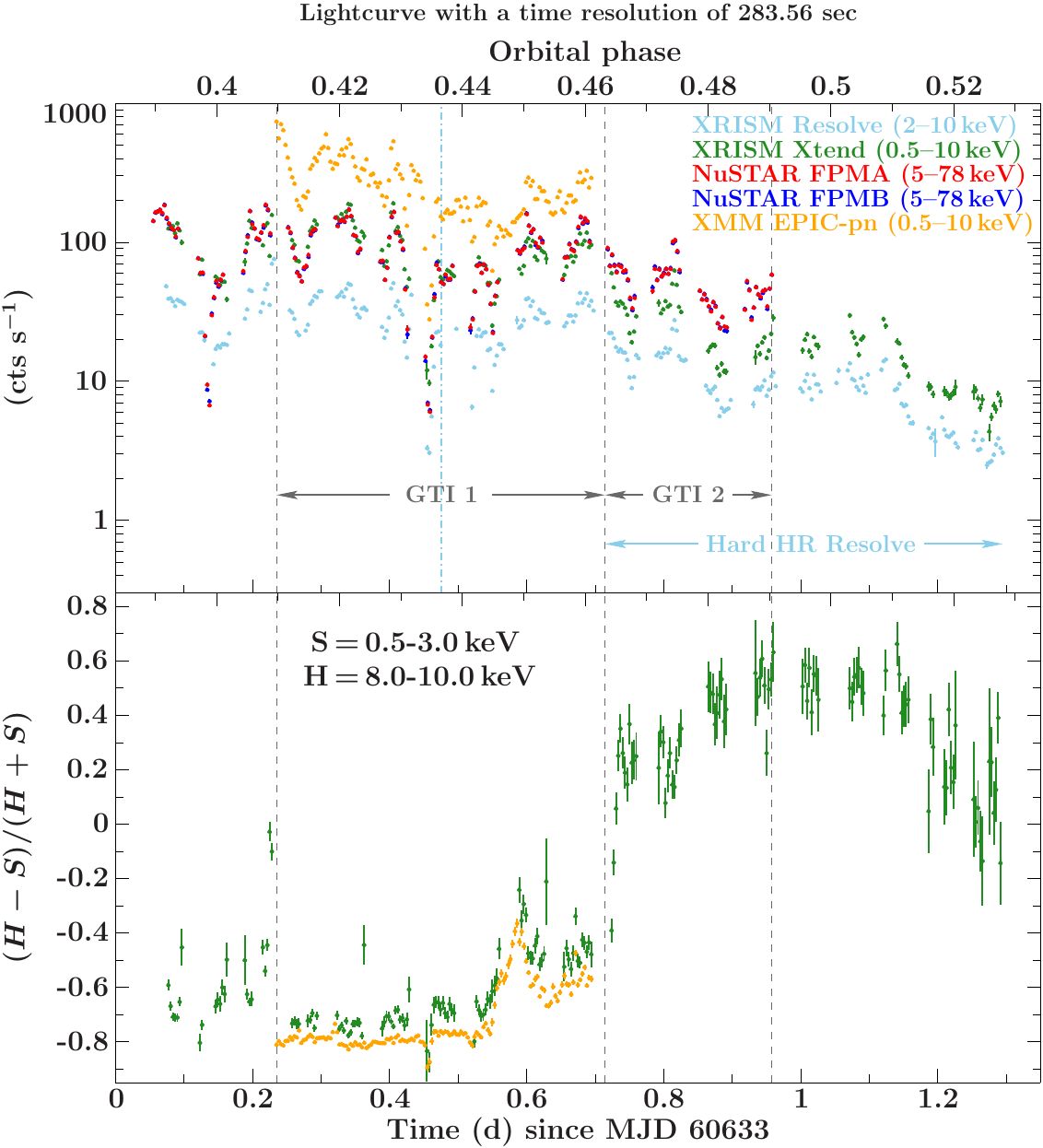}}
    \caption{Light curves and hardness ratio for \xrism/Resolve (light blue) and Xtend (green), \xmmnewton/EPIC-pn (orange), and \nustar/FPMA (red) and FPMB (blue) with a time resolution of $P = 283.595\,\rm{s}$. \textit{Top panel:} Overall count rates in the energy bands corresponding to the respective instruments. The black dashed vertical lines denote the two GTIs we chose for our spectral analysis (Sect.~\ref{subsection:low_HR} and Sect.~\ref{subsection:high_HR}). The dash-dotted light blue line marks the end of the FW electronics error, i.e. the beginning of the well-calibrated \xrism/Resolve data for our spectral analysis (Sect.~\ref{subsection:high_HR_rsl}). \textit{Bottom panel:} Hardness ratio between the 0.5--3\,keV (soft) and 8--10\,keV (hard) energy bands.}
    \label{fig:lc_hr}
\end{figure}

Vela X-1 is known for its strong flux variability on multiple timescales, from kiloseconds to individual neutron star pulses \citep[see e.g.][]{Martinez_2014, Diez_2022, Diez_2023}. To examine the source's variability during our observation, we present in Fig.~\ref{fig:lc_hr} the 0.5--10\,keV \xmmnewton/EPIC-pn, 5--79\,keV \nustar, 2--10\,keV \xrism/Resolve, and 0.5--10\,keV \xrism/Xtend light curves. The pulse period ($P$) of Vela X-1 was independently determined using the $Z^2$ statistics~\citep{Buccheri_1983_Z2}, based on photon events. We collected the photon events from all the instruments and corrected them for both barycentric motion and binary orbital effects using the orbital solution set by the Fermi Gamma-ray Burst Monitor Accreting Pulsars Program \citep{Malacaria_2020}. The results are consistent, yielding a period of $P = 283.595 \pm 0.025\,\mathrm{s}$. The light curves shown in Fig.~\ref{fig:lc_hr} were binned accordingly to mitigate flux variations during one pulse cycle. The same Figure also shows the hardness ratio (HR) between the $0.5$--$3.0$\,keV and $8.0$--$10.0$\,keV energy bands. Two distinct phases are apparent in the HR: a soft HR interval and a hard HR interval. The HR serves as a reliable proxy for the absorption level in sources of this type \citep[e.g.][]{Abalo_2024}. In particular, a sharp transition from soft to hard HR was observed at $\phi_{\mathrm{orb}} \approx 0.463$ in Fig.~\ref{fig:lc_hr}, which is typically attributed to the onset of accretion and ionisation wakes entering our line of sight, leading to increased absorption and spectral hardening \citep[see detailed analysis in][]{Diez_2023}. 

\subsection{Soft HR (GTI~1): \xmmnewton + \nustar + \xrism}
\label{subsection:low_HR}

For the soft HR interval, we selected the good time interval (GTI) during which both low- and high-energy data are simultaneous. This corresponds to the interval from $\sim$60633.23\,MJD ($\phi_{\mathrm{orb}} \approx 0.410$) to $\sim$60633.71\,MJD ($\phi_{\mathrm{orb}} \approx 0.463$), referred to as `GTI~1' in Fig.~\ref{fig:lc_hr}. This selection ensured constraints on the high-energy continuum while allowing us to investigate absorption variability at lower energies. Due to the filter wheel (FW) electronics error that compromised \xrism/Resolve data before $\sim$60633.47\,MJD (see Fig.~\ref{fig:lc_hr} and Sect.~\ref{appendix:data_reduc}), we excluded Resolve data before that MJD for this analysis. Since the \xmmnewton/EPIC‑pn CCD camera already covers the 0.5–10\,keV band and \xrism/Xtend operated with only $\sim$12\% live-time (see Table~\ref{tab:obs_details}), we excluded Xtend from our analysis. This choice simplifies the spectral modelling by avoiding additional cross-calibration uncertainties beyond those already documented between \xmmnewton\ and \nustar\ \citep[see Appendix in][]{Diez_2023}. These will be addressed in a future study that is currently in preparation. 

To maintain consistency with previous studies \citep[][]{Fuerst_2014, Diez_2022, Diez_2023}, we adopted a partial covering model modified by a Fermi-Dirac high-energy cut-off \citep[\texttt{FDcut};][]{Tanaka_1986} that can be written as 
\begin{align}
\label{eq:final_model}
\begin{split}
\mathcal{M}(E) =& \ \texttt{Tbabs}_{\mathrm{ISM}} \times \texttt{Tbpcf} \ \times \\ &(\texttt{Powerlaw} \times \texttt{FDcut} \times \prod^3_{i=1}\texttt{Gabs}_i \ + \sum^{n}_{j=1}\texttt{Gauss}_j)
\end{split}
\end{align}
where: $\texttt{Powerlaw} = K\,E^{-\Gamma}, \ \texttt{FDcut} = \left(1+ \exp \left(\frac{E-E_{\rm{cut}}}{E_{\rm{fold}}} \right) \right) ^{-1}$, \\ $\texttt{Tbabs}_{\mathrm{ISM}} = \exp^{-N_{\rm{H,ISM}}\,\sigma(E)}, \texttt{Tbpcf} = {\rm{CF}} \times \exp^{-N_{\rm{H}}\,\sigma(E)} +\,(1-{\rm{CF}})$. 

Here, $N_{\rm{H}}$ is left free to vary and represents the equivalent hydrogen column density of the dense stellar wind surrounding the neutron star. The interstellar medium (ISM) column density, $N_{\rm{H,ISM}}$, is fixed at $3.71 \times 10^{21}\,\rm{cm^{-2}}$ based on the NASA HEASARC $N_{\rm{H}}$ tool \citep{HI4PI_2016}. Abundances are from \cite{Wilms_2000}, and the cross-section $\sigma$ is from \cite{Verner_1996}. The covering fraction (CF), ranging from zero to one, quantifies the opacity of the obscuring material. To correct for known calibration uncertainties in the EPIC-pn timing mode \citep[e.g.][]{Diez_2023, Lai_2024} and additional discrepancies with \xrism/Resolve, we introduced two cross-calibration normalisation constants, $\mathcal{C}_{\mathrm{EPIC-pn}}$ and $\mathcal{C}_{\mathrm{Resolve}}$. We also accounted for discrepancies between \nustar's FPMA and FPMB with $\mathcal{C}_{\mathrm{FPMB}}$, setting FPMA as the reference for all detectors. 

Three multiplicative broad Gaussian absorption components (\texttt{Gabs}) were used to model the two cyclotron resonance scattering features (CRSFs) and the 10-keV absorption feature. For additional emission or absorption lines, we used a phenomenological approach by fitting Gaussian components when significant residuals are present. Instead of judging the fit only in the sample or relying on asymptotic penalties (e.g. Akaike Information Criterion), we compared the expected log pointwise predictive density (elpd), as implemented in \texttt{ArviZ} \citep{Vehtari_2015}, between models with and without line components. For this GTI~1, we detected five emission lines. Three lines are associated with the fluorescent Fe K$\alpha_2$, Fe K$\alpha_1$, and Fe K$\beta$ transitions at around 6.39\,keV, 6.40\,keV, and 7.06\,keV, respectively \citep{Holzer_1997}. These lines arise from photoionisation of Fe atoms with subsequent transitions from the L-shell and M-shell down to the K-shell producing the K$\alpha$ doublet and the K$\beta$ line, respectively. We also observed residuals around 1\,keV and 1.3\,keV corresponding to unresolved blended ionised lines in the Ne and Mg regions, which are typically associated with the presence of an ionised plasma around the X-ray source \citep[e.g.][]{Amato_2021}. We display in Fig.~\ref{fig:low_abs_spec} in the Appendix the simultaneous \nustar, \xmmnewton/EPIC-pn, and \xrism/Resolve spectra along with our best-fit model. We report in Table~\ref{tab:best_fit_params} in the Appendix the corresponding parameter values. Although we can place upper limits on the normalisation and width of the fluorescent Ni K$\alpha$ transition at $\sim$7.45 keV \citep{Holzer_1997}, the values are mostly consistent with zero. Moreover, including the Ni K$\alpha$ line emission does not improve the elpd sufficiently to claim a detection (see inset of Fig.~\ref{fig:low_abs_spec}). 

Focusing on the stellar wind parameters obtained during this GTI~1, the covering fraction CF is $0.912^{+0.004}_{-0.003}$, and the equivalent hydrogen absorption column density of the stellar wind, $N_{\mathrm{H}}$, is $(7.46^{+0.18}_{-0.17}) \ \times 10^{22} \ \mathrm{cm^{-2}}$. These values are consistent with the previous results reported by \citet{Diez_2023}, which were also obtained at a similar $\phi_{\rm{orb}}$ ($\sim$0.5) and soft HR interval. 

We note that physically motivated reflection models have also been used to describe the complex wind geometry of Vela~X-1, such as \texttt{pexrav} in \citet{Rahin_2023} or \texttt{MYTORUS} in \citet{Tzanavaris_2018}. However, \citet{Rahin_2023} argued that reflection from the star's photosphere alone is not sufficient to explain discrepancies they observed at $\phi_{\rm{orb}} \approx 0.3$--$0.7$ using \nicer observations and that additional reflection may originate from clumps and from the accretion stream at these orbital phases. Additionally, \citet{Tzanavaris_2018} conclude that the spectrum of the \chandra observation they analysed at $\phi_{\rm{orb}} \approx 0.5$ is no longer reflection dominated, in contrast to another observation taken during the eclipse. A thorough investigation of the aforementioned models along with other physically motivated models for accreting X-ray pulsars available in the literature \citep[e.g. \texttt{compmag} in][]{Farinelli_2016} in the context of our dataset is beyond the scope of this work. We note that our phenomenological approach employing a partial covering model supports the presence of a non-uniform and complex wind geometry, which is consistent with most of the literature, regardless of whether the data were modelled using physical or phenomenological approaches.

\subsection{Hard HR (GTI~2): \xrism + \nustar}
\label{subsection:high_HR}

As in Sect.~\ref{subsection:low_HR}, we selected low- and high-energy simultaneous events during the hard HR interval, excluding \xrism/Xtend. For this GTI, no \xmmnewton coverage is available. Hence, we considered simultaneous \nustar and \xrism/Resolve data between $\sim$60633.71\,MJD ($\phi_{\mathrm{orb}} \approx 0.463$) and $\sim$60633.96\,MJD ($\phi_{\mathrm{orb}} \approx 0.490$), referred to as `GTI~2' in Fig.~\ref{fig:lc_hr}. 

We show in Fig.~\ref{fig:high_abs_spec} in the Appendix the \nustar and \xrism/Resolve simultaneous spectra with our best-fit model. We list in Table~\ref{tab:best_fit_params} the corresponding parameter values. In this GTI~2, we detected three emission lines corresponding again to the Fe K$\alpha_2$, Fe K$\alpha_1$, and Fe K$\beta$ fluorescent lines. As in GTI~1, the normalisation and width values of the Ni K$\alpha$ line are consistent with zero with no strong detection (see inset of Fig.~\ref{fig:high_abs_spec}). The previously detected soft lines below 2\,keV in GTI~1 are not visible in this GTI~2. This is a consequence of the Resolve's closed GV, the absence of \xmmnewton coverage, and our decision to exclude Xtend due to drastically reduced exposure time. 

Focusing on the stellar wind influence during this GTI~2, we obtained a covering fraction CF of $0.882\pm0.008$ and an equivalent hydrogen absorption column density of the stellar wind $N_{\mathrm{H}}$ of $(25.04^{+1.00}_{-0.96}) \ \times 10^{22} \ \mathrm{cm^{-2}}$. As in GTI~1, the CF remains close to one, indicating that almost 100\% of the line of sight is obscured when the neutron star is near inferior conjunction. However, during GTI~2, the stellar wind $N_{\mathrm{H}}$ is approximately three times higher than in GTI~1, which is consistent with the observed hardening of the spectral shape (Fig.~\ref{fig:lc_hr}) and likely due to increased absorption as reported by \citet{Abalo_2024} with 14 years of MAXI data. This high absorption is also evidenced by the presence of the Fe K-edge at $\sim$7.1 keV in the Resolve spectrum, a feature that becomes more prominent with a high $N_{\mathrm{H}}$ \citep[see Fig.~3 in][]{Diez_2023}.

\subsection{Hard HR \xrism averaged spectrum}
\label{subsection:high_HR_rsl}

\begin{figure*}
    \centering
    \centerline{\includegraphics[trim=0cm 0cm 0cm 0cm, clip=true, width=1.0\linewidth]{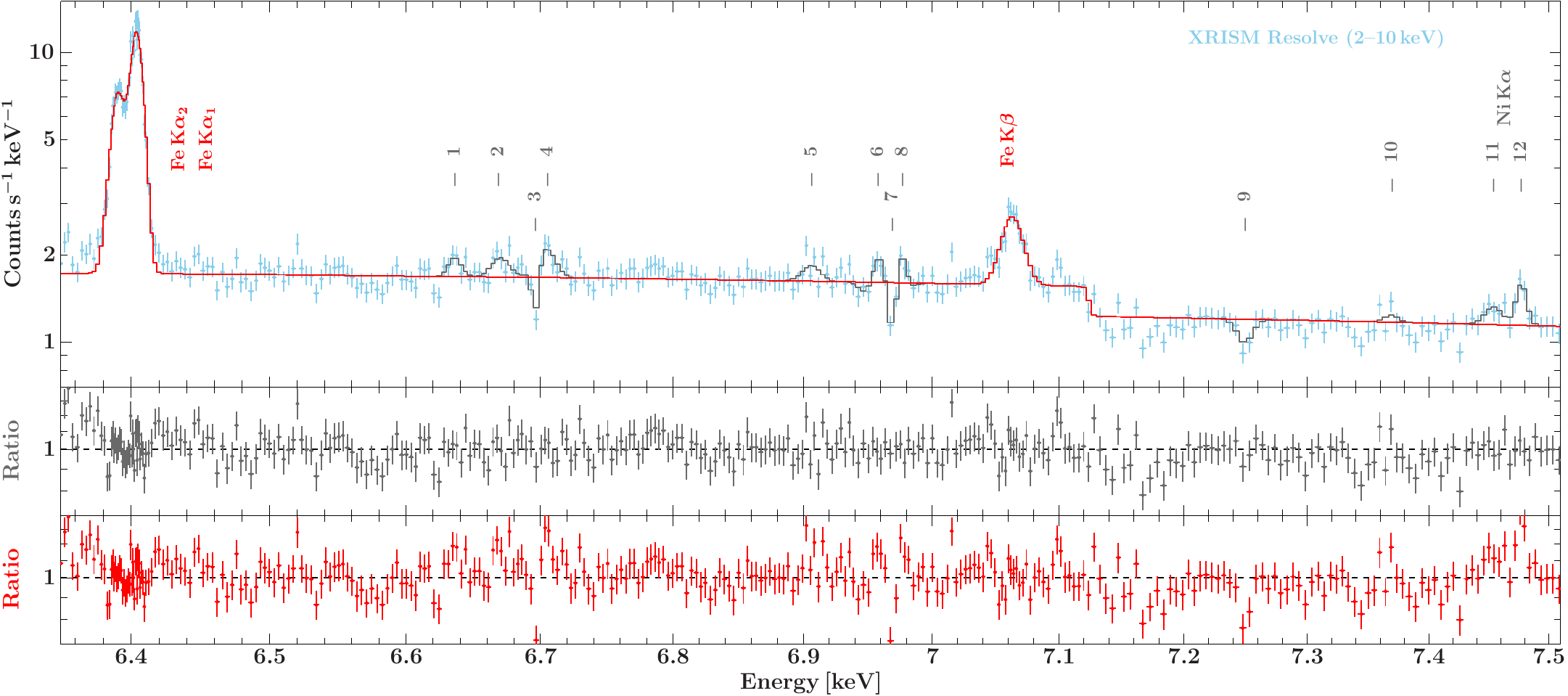}}
    \caption{Folded \xrism/Resolve spectrum (light blue) averaged over the hard HR interval. The red solid line corresponds to the base model we used to test for the presence of additional line components, indicated by the grey solid line (see Sect.~\ref{subsubsection:lines}). Lines for which a detection could be resolved are labelled by their names; the others are labelled by numbers. The corresponding ratio residuals in the lower panels are computed as data/model. For plotting purposes, the data were binned to a minimum of 150 counts/bin.}
    \label{fig:all_lines_spec}
\end{figure*}

In order to have a look at the full \xrism/Resolve spectrum during the highly absorbed phase without constraints of GTI selection (Fig.~\ref{fig:lc_hr}), we show in Fig.~\ref{fig:all_lines_spec} the \xrism/Resolve spectrum averaged over the full hard HR interval. This stable and long time interval mitigates variations caused by the high variability of the source and enhances the visibility of ionised lines in the spectrum thanks to a higher absorption. Hence, it allows for a deeper and pertinent investigation of the presence of line components in the Fe region. The continuum model we used here is our best-fit model from Eq.~\ref{eq:final_model} with the parameters not covered by Resolve removed, such as the \texttt{FDcut} and CRSFs. We present the best-fit parameters in Table~\ref{tab:best_fit_params} in the Appendix, and our search for line components is described in the following subsections.

\subsubsection{Search for a Compton shoulder}
\label{subsubsection:CS}

When investigating the presence of a Compton shoulder (CS), we started by looking in the Fe lines, a feature that can be observed in HMXBs and that shows a strong Fe K$\alpha$ complex \citep[see e.g.][for GX 301--2]{Watanabe_2003}. When a high-energy photon propagates through matter, it has a non-negligible probability of interacting with electrons via Compton scattering. In this process, the photon loses energy (down-scattering), and if the original photon belongs to a strong emission line, this can produce a discernible Compton shoulder on the low-energy side of the observed line. We thus tied the energies of two Gaussian components to the Fe K$\alpha_1$ and K$\alpha_2$ line energies with a 156 eV offset as expected from the Compton formula \citep{Watanabe_2003}. When comparing the elpd between models with and without the tied Gaussians, we did not observe any significant improvement of the fit with the addition of the CS, and the upper normalisation values are consistent with zero. We also tested a skewed Gaussian for the Fe K$\alpha_2$ line in case of a very weak CS, but the results led to the same conclusion. Hence, we did not include a CS component in our model.

\subsubsection{Line detections in the Fe region}
\label{subsubsection:lines}

\begin{figure}[htpb]
    \centering
    \centerline{\includegraphics[trim=0cm 0cm 0cm 0cm, clip=true, width=1\linewidth]{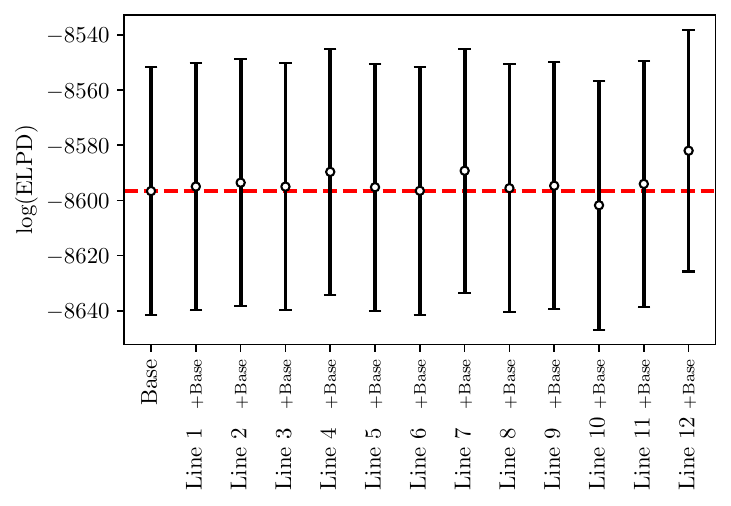}}
    \caption{Comparison of the expected log predictive densities between the base model and the same model adding one line component at a time. The base model includes the Fe K$\alpha_1$, Fe K$\alpha_2$, and Fe K$\beta$ lines.}
    \label{fig:elpd}
\end{figure}

We also searched for the presence of other emission or absorption line components in the Fe region. We fit Gaussian components with centroids close to reference energies of known transitions based on the database compiled by \citet{hell_2025} and where residuals are observed in the spectrum. The identification of spectral lines is uncertain and often ambiguous since many observed features could result from blends of unresolved fine-structure transitions or overlapping contributions from multiple ionisation states. The line components we tested are indicated in Fig.~\ref{fig:all_lines_spec} by the grey line and corresponding numbers. In Fig.~\ref{fig:elpd}, we compare the elpd between the `base' model, which consists of our best-fit continuum and includes previous confidently detected lines in this work (Fe K$\alpha_1$, Fe K$\alpha_2$, and Fe K$\beta$), and the same model adding one line at a time as labelled in Figs.~\ref{fig:all_lines_spec} and \ref{fig:elpd}. We observed that adding Line 12, which coincides well with the fluorescent Ni~K$\alpha$ complex, yielded the largest improvement to the base model. In contrast to GTI~1 and GTI~2, this time we could constrain the Ni K$\alpha$ parameters as shown in Table~\ref{tab:best_fit_params}. Nonetheless, this weak detection fails to yield a statistically significant improvement in the fit once uncertainties are accounted for, as can be seen in Fig.~\ref{fig:elpd}. Similarly, we cannot confidently claim any additional line detections in our dataset because of the low signal-to-noise ratio. This may result from a combination of a reduced exposure time due to the FW electronics error and a difference in orbital phase: the ER covered $\phi_{\mathrm{orb}} \approx 0.68$--$0.90$ (see Fig.~\ref{fig:orbit}), where higher absorption of $\sim$50$\times 10^{22}\,{\mathrm{cm^{-2}}}$ is expected \citep{Diez_2022}. At such phases, the ionisation wake becomes more prominent in our line of sight, enhancing the visibility of ionised lines in the spectrum \citep{Diez_2023}, which may lead to a stronger ionised Fe XXV complex around 6.6--6.7\,keV, as seen in the \xrism ER data. 

If present, lines 1--2--3--4 coincide well with the He-like Fe~XXV complex (w, x, y, z), and lines 6--7--8 coincide well with the H-like Fe XXVI complex (Ly$\alpha_{1-2-3}$) \citep{Yerokhin_2019, Yerokhin_2015}. These emission lines are accompanied by absorption lines that could indicate a P-Cygni profile. Lines 11--12 likely correspond to the fluorescent Ni K$\alpha_2$ and K$\alpha_1$ transitions \citep{Holzer_1997} in the Ni K$\alpha$ complex. For lines 5--9--10, the identification is more uncertain. Although their energies match with Co K$\alpha$, Co XXVI w, and Cr XXIII w$_7$, respectively \citep{Holzer_1997, Yerokhin_2019}, the negligible abundances of Co and Cr compared to Fe, both in solar and ISM abundances \citep{Wilms_2000}, make this interpretation very unlikely. These features could also result from blue- or red-shifted transitions of other elements. Due to insufficient statistics to confirm clear detections, we did not investigate the physical properties of these lines in this dataset further.

Looking at Table~\ref{tab:best_fit_params}, we noticed, as in GTI~2, that the energies of the fluorescent lines in the Fe region are found at their laboratory measurements \citep{Holzer_1997}, within the uncertainties. Hence, the observed broadening is likely due to turbulent motions in the wind and the intrinsic line shape for which the velocities can be derived by $\sigma_{\mathrm{line}}/{E_{\mathrm{line}}} \approx v/c$ for non-relativistic motions ($v\ll c$) and computed using the posterior samples from the variational inference. We found $193^{+17}_{-18} \ \mathrm{km\,s^{-1}}$, $225^{+37}_{-36} \ \mathrm{km\,s^{-1}}$, $341^{+58}_{-63} \ \mathrm{km\,s^{-1}}$, and $146^{+114}_{-106} \ \mathrm{km\,s^{-1}}$ for the Fe K$\alpha_1$, Fe K$\alpha_2$, Fe K$\beta$, and Ni K$\alpha$ lines, respectively. 

\section{Discussion and conclusion}
\label{section:conclusion}

The high resolving power of \xrism/Resolve provides new insights into the spectra of wind-accreting HMXBs, in particular their Fe region. Here, we have presented the results from a \xrism observation of Vela X-1 during the first cycle of the general observer programme complemented by simultaneous \xmmnewton and \nustar coverage. The line energies of the Fe K$\alpha_1$ and Fe K$\alpha_2$ doublet are consistent with near-neutral or lowly ionised ions \citep[Fe\,{\sc xii} \ or less,][]{Kallman_2004}. Hence, observing strong fluorescent lines points towards a low to moderate ionisation state of the material despite an intense X-ray photoionisation, which is likely attributable to cold and dense clumps embedded in a hot and photoionised wind \citep{Sako_1999, Amato_2021}. In \chandra spectra at $\phi_{\mathrm{orb}} \approx 0.5$, \citet{Goldstein_2004} detected the additional fluorescent lines, such as Ni K$\alpha$ and Fe K$\beta$, that we also observed in our \xrism/Resolve spectrum at the same orbital phase. We measured the wind velocities to be of the order of $\sim$$10^2\,\mathrm{km\,s^{-1}}$ using the Fe K$\alpha_1$, Fe K$\alpha_2$, Fe K$\beta$, and Ni K$\alpha$ lined broadening. These values are consistent with theoretical estimates of wind velocities in the inner region where the neutron star is located, according to \cite{Sander_2018}, who found $v(d_{\mathrm{NS}}) \approx 100\,\mathrm{km\,s^{-1}}$. The velocity perturbations observed in the line-driven wind of hot stars create instabilities and subsequently turn into dense regions of gas, which explains the presence of clumps \citep[see the review by][]{Martinez_2017}. Therefore, these near-neutral fluorescent lines were likely produced in cool clumps of gas in the vicinity of the neutron star during this observation, although \cite{Sato_1986} and \cite{Watanabe_2006} have also reported the presence of these lines further in the accretion wake with observations during the eclipse. 

In this study, we have also demonstrated the pertinence and effectiveness of variational inference using the \texttt{jaxspec} X-ray spectral fitting library to produce meaningful constraints on the high-resolution spectra produced by \xrism/Resolve. All of the parameters of the fits were left to vary thanks to the use of variational inference with a multivariate Gaussian guide for the posterior parameters, which was not possible within the classical framework of \texttt{ISIS} for some parameters such as $E_{\rm{cut}}$ or $E_{\rm{fold}}$. 

We detected a sudden increase in the HR, a behaviour already seen in Vela X-1 when the neutron star approaches an inferior conjunction around $\phi_{\mathrm{orb}} \approx 0.5$ \citep{Diez_2023}, which is attributed to the onset of the wakes crossing our line of sight. This transition occurs slightly earlier than in previous observations \citep{Diez_2023}, confirming that the structures causing absorption variations are not stable in orbital phase. Future time-resolved spectroscopy tracking the evolution of the absorption column density, $N_{\mathrm{H}}$, will help constrain the presence and properties (e.g. mass, density) of clumps in the stellar wind. In this study, we have analysed only 79.3\% of the total available \nustar exposure. We have not yet investigated the $\sim$6-ks \xrism/Xtend or the $\sim$42-ks \xmmnewton/RGS datasets, which will be addressed in a forthcoming publication.

\begin{acknowledgements}
We thank the anonymous referee for their valuable feedback which improved the quality of this work. CMD and EQ acknowledge support through the European Space Agency (ESA) Research Fellowship Programme in Space Science. SD acknowledges the support of CNRS/INSU and CNES. This project was provided with HPC and storage resources by GENCI at IDRIS thanks to the grant 2024-AD010416032 on the supercomputer Jean Zay's A100/ H100 partitions. This work has made use of (1) the Interactive Spectral Interpretation System (\texttt{ISIS}) maintained by Chandra X-ray Center group at MIT; (2) the \nustar Data Analysis Software (NuSTARDAS) jointly developed by the ASI Science Data Center (ASDC, Italy) and the California Institute of Technology (Caltech, USA); (3) the \texttt{ISIS} functions (\texttt{isisscripts})\footnote{\url{http://www.sternwarte.uni-erlangen.de/isis/}\label{footnote:isisscripts}} provided by ECAP/Remeis observatory and MIT; (4) NASA's Astrophysics Data System Bibliographic Service (ADS). We thank John E. Davis for the development of the \texttt{slxfig}\footnote{\url{http://www.jedsoft.org/fun/slxfig/}\label{footnote:jedsoft}} module used to prepare most of the figures in this work. This work made use of the \texttt{JAX} \citep{bradbury_jax_2018} and \texttt{numpyro} \citep{phan_composable_2019} packages.

\end{acknowledgements}

\bibliographystyle{aa}
\bibliography{references}

\begin{appendix}

\section{Data reduction}
\label{appendix:data_reduc}

\begin{table}[!h]
\caption{Observations log.}
\label{tab:obs_details}
\begin{center}
\begin{small}
\begin{tabular}{ccc}    
\hline\hline
Mission + Instrument &
Obs ID & 
Exposure (ks)\\
\hline
\xrism (Resolve/Xtend) & {201131010}     & 59.295$^{(a)}$/5.992$^{(b)}$    \\
\xmmnewton (EPIC-pn) & {0953790501}     &  39.145  \\
\nustar (FPMA/B) & {91002349002}     & 36.932/37.083   \\
\hline
\multicolumn{3}{p{0.9\linewidth}}{$^{(a)}$ Reduced to 39.034 ks of usable data for the spectral analysis due to the FW electronics error (see Sect.~\ref{appendix:data_reduc}).}\\
\multicolumn{3}{p{0.9\linewidth}}{$^{(b)}$ Xtend was set in 1/8 window + 0.1 sec burst mode which results in a live time fraction of about 12\% (XRISM Proposers' Observatory Guide Section 6.2).}\\
\end{tabular}
\end{small}
\end{center}
\end{table}

The EPIC-pn observation is significantly affected by pile-up. To mitigate this, we use the \texttt{epatplot} tool, following the approach outlined in the Appendix of \citet{Diez_2023}. Specifically, we exclude the three centremost columns of the point spread function (PSF), \texttt{RAWX 36--37--38}, which are most severely impacted. A Jupyter Notebook tutorial for the \xmmnewton\ data extraction of another observation of Vela X-1 is provided in \citet{Guelbahar_2025} and follows a procedure similar to the one used here. As reported in \citet{Diez_2023}, we also observe count-rate-dependent energy scale shifts in the EPIC-pn spectrum of Vela X-1, as evidenced by the Fe K$\alpha$ line appearing at $\sim$6.56\,keV instead of the expected 6.4\,keV. Since the release of SAS v22.1.0, an updated version of the \texttt{evenergyshift}\footnote{\url{https://www.cosmos.esa.int/web/xmm-newton/sas-thread-evenergyshift}} task has been made available\footnote{\url{https://www.cosmos.esa.int/web/xmm-newton/sas-release-notes-2210}}. Following the recommended procedure, we measured energy shifts in the time-averaged spectrum near the instrumental edges and applied the \texttt{evenergyshift} correction, which resulted in the Fe K$\alpha$ line shifting to $\sim$6.42\,keV. We did not need to filter for flaring particle background or extract a background, because the source was so bright that it illuminated the entire EPIC-pn focal plane.

Resolve is a soft X-ray spectrometer composed of a $6 \times 6$ array of microcalorimeters with an excellent spectral resolution of 4.5\,eV FWHM at 6\,keV \citep{Tashiro_2025}. Unfortunately, the X-ray aperture door -- known as the gate valve (GV) -- remains closed despite several attempts to open it, effectively blocking soft X-rays below 2\,keV and attenuating the overall bandpass. In addition, during our observation, a filter wheel (FW) electronics error (now resolved) prevented the planned rotation of the $^{55}$Fe FW calibration source \citep[for more details, see][]{Porter_2016,deVries_2018}, compromising the spectral data quality before $\sim$60633.47\,MJD. Nevertheless, thanks to the high count rate of Vela X-1, we have sufficient photon statistics to meet most of our scientific objectives.

The orbital phases $\phi_{\mathrm{orb}}$ mentioned in this work are derived from the ephemeris of \citet{Bildsten_1997} and \citet{Kreykenbohm_2008}, as summarised in Table~1 of \citet{Diez_2022}, where $\phi_{\mathrm{orb}} = 0$ is defined as the time when the mean longitude $l$ equals 90° ($T_{90}$), following the convention commonly used for this source. All times are barycentred and corrected for the binary orbit.  All spectra are rebinned using the optimal approach from \cite{Kaastra_2016}, with an additional criterion of a minimum of 50 cts/bin for \xrism/Resolve due to the high statistics that is enabled by the instrument. Errors are given within 2-$\sigma$ confidence level, unless stated otherwise. We use HEASOFT v6.35.1 for the reduction of data from the NASA-led missions \nustar\ and \xrism. For \xmmnewton\ data extraction, we use the Science Analysis System (SAS) software v22.1.0.

\section{Spectra for GTI~1 and GTI~2}
\label{appendix:GTI12}

\begin{figure}[htpb]
    \centering
    \centerline{\includegraphics[trim=0cm 0cm 0cm 0cm, clip=true, width=1.0\linewidth]{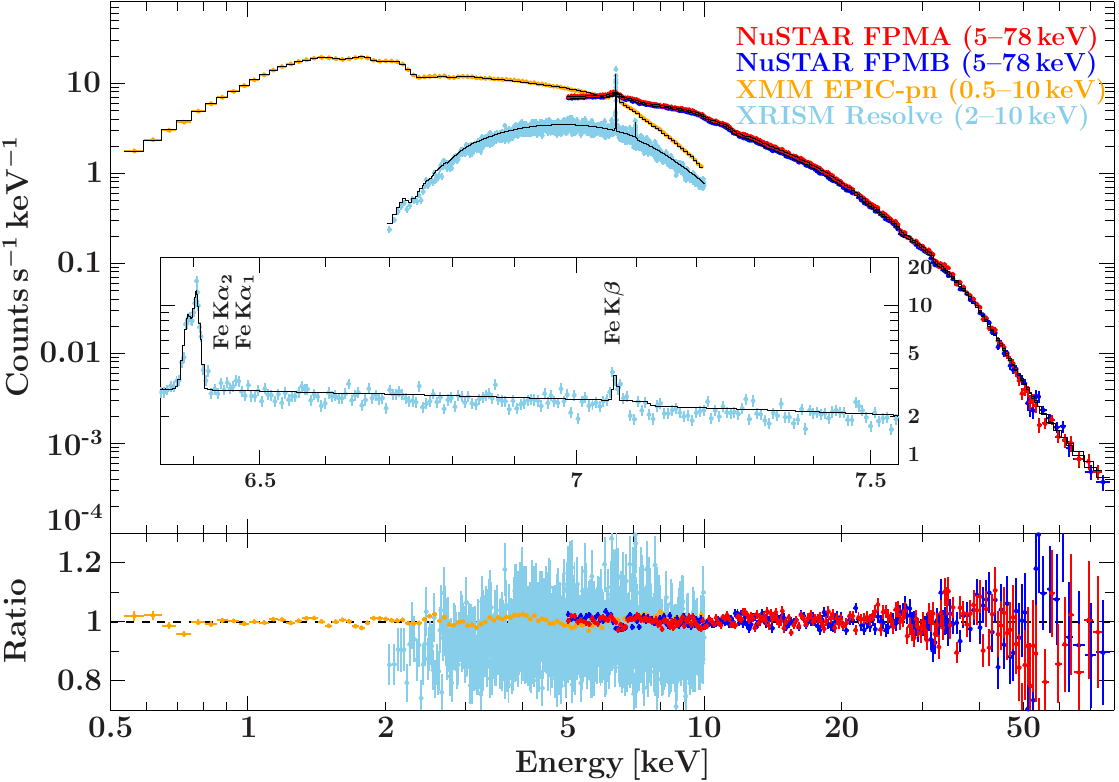}}
    \caption{Folded spectra for GTI~1 obtained with \xmmnewton/EPIC-pn (orange), \nustar/FPMA (red) and FPMB (dark blue), and \xrism/Resolve (light blue). Solid lines show the best-fit model and corresponding ratio in the bottom panel computed as data/model. A zoom in the Fe region is shown to highlight the resolution capabilities of Resolve. For plotting and readability purposes, the \nustar and \xmmnewton data are not displayed in the zoomed window and the \xrism/Resolve data are binned to a minimum of 150 counts/bin.}
    \label{fig:low_abs_spec}
\end{figure}

\begin{figure}[htpb]
    \centering
    \centerline{\includegraphics[trim=0cm 0cm 0cm 0cm, clip=true, width=1.0\linewidth]{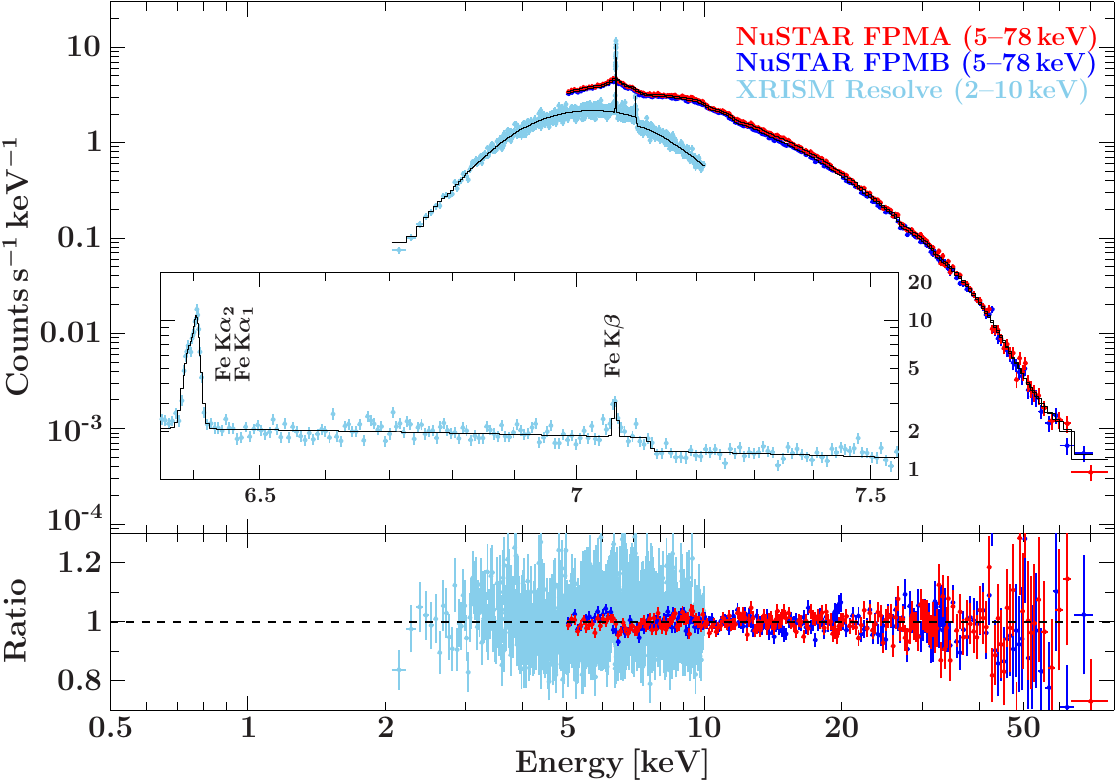}}
    \caption{Folded spectra for GTI~2 obtained with \xrism/Resolve (light blue), and \nustar/FPMA (red) and FPMB (dark blue). Solid lines show the best-fit model and corresponding ratio in the bottom panel computed as data/model. A zoom in the Fe region is shown to highlight the resolution capabilities of Resolve. For plotting and readability purposes, the \nustar data are not displayed in the zoomed window and the \xrism/Resolve data are binned to a minimum of 150 counts/bin.}
    \label{fig:high_abs_spec}
\end{figure}

\onecolumn

\newpage

\section{Best-fit parameters}
\label{tab:best_fit_params}

\begin{table}[!htpb]
\begin{center}
\begin{small}
\begin{tabular}{llll}    
\hline
\multicolumn{1}{l}{Parameter} & 
\multicolumn{1}{c}{GTI~1 (soft HR)} &
\multicolumn{1}{c}{GTI~2 (hard HR)} &
\multicolumn{1}{c}{Hard HR average spectrum} \\
&
\multicolumn{1}{c}{\nustar + \xmmnewton + \xrism} &
\multicolumn{1}{c}{\nustar + \xrism} &
\multicolumn{1}{c}{\xrism} \\
\hline

$C_{\rm{FPMA}}$  &  fixed to 1  &  fixed to 1 & -- \\
$C_{\rm{FPMB}}$  & $1.011\pm0.003$ & $1.010\pm0.005$ & -- \\
$C_{\rm{EPIC-pn}}$  &  $0.879\pm0.003$  &  -- & -- \\
$C_{\rm{Resolve}}$  &  $0.943\pm0.005$  &  $1.003^{+0.007}_{-0.007}$ & fixed to 1 \\
$N_{\rm{H,ISM}} \ (10^{22} \, \rm{cm^{-2}})$  & fixed to 0.371 & fixed to 0.371 & fixed to 0.371\\
$N_{\rm{H}} \ (10^{22} \,  \rm{cm^{-2}})$  &  $7.46^{+0.18}_{-0.17}$ & $25.04^{+1.00}_{-0.96}$ & $31.15^{+1.10}_{-1.07}$ \\
$\rm{CF}$  & $0.912^{+0.004}_{-0.003}$ & $0.882\pm0.008$ & $0.876\pm0.007$ \\
$\Gamma$  & $0.91\pm0.01$ & $1.15\pm0.02$ & $0.90\pm0.04$ \\
$K \ (\rm{ph \ keV^{-1} \ cm^{-2} \ s^{-1}})$  &  $0.322\pm0.011$  &  $0.245\pm0.013$ & $0.096\pm0.007$ \\
$E_{\rm{cut}} \ (\rm{keV})$  &  $19.87^{+1.07}_{-1.13}$ & $20.02^{+1.47}_{-1.39}$ & -- \\
$E_{\rm{fold}} \ (\rm{keV})$  &  $13.12^{+0.46}_{-0.42}$ & $13.14^{+0.38}_{-0.35}$  & -- \\
$E_{10\,\rm{keV}} \ (\rm{keV})$  &  $9.5\pm0.2$ & $10.3\pm0.2$ & $9.3\pm0.2$ \\
$d_{10\,\rm{keV}} \ (\rm{keV})$  &  $4.63^{+0.26}_{-0.25}$ & $0.69^{+0.12}_{-0.13}$ & $0.87^{+0.20}_{-0.21}$ \\
$\sigma_{10\,\rm{keV}} \ (\rm{keV})$  &  $4.85^{+0.26}_{-0.24}$ & $2.77\pm0.29$ & $1.42^{+0.21}_{-0.21}$ \\
$E_{\rm{CRSF,F}} \ (\rm{keV})$  &  $23.01^{+0.30}_{-0.32}$ & $24.34^{+0.82}_{-0.88}$ & -- \\
$\sigma_{\rm{CRSF,F}} \ (\rm{keV})$  & $0.5\times \sigma_{\rm{CRSF,H}}$ & $0.5\times \sigma_{\rm{CRSF,H}}$ & -- \\
$d_{\rm{CRSF,F}} \ (\rm{keV})$  &  $8.92^{+0.41}_{-0.40}$ & $0.84^{+0.27}_{-0.28}$ & -- \\
$E_{\rm{CRSF,H}} \ (\rm{keV})$  &  $55.67^{+0.62}_{-0.63}$ & $53.14^{+0.97}_{-0.87}$ & -- \\
$\sigma_{\rm{CRSF,H}} \ (\rm{keV})$  &  $14.79^{+0.38}_{-0.40}$ & $8.41^{+0.79}_{-0.77}$ & -- \\
$d_{\rm{CRSF,H}} \ (\rm{keV})$  &  $68.3^{+5.0}_{-4.8}$ & $17.5^{+1.9}_{-2.1}$ & -- \\
$E_{{\mathrm{Ne}}} \ (\rm{keV})$  &  $1.18\pm0.01$ & -- & -- \\
$A_{{\mathrm{Ne}}} \ (\rm{ph \, s^{-1} \, cm^{-2}})$  &  $0.357^{+0.015}_{-0.018}$ & -- & -- \\
$\sigma_{{\mathrm{Ne}}} \ (\rm{keV})$  &  $0.993^{+0.005}_{-0.007}$ & -- & -- \\
$E_{{\mathrm{Mg}}} \ (\rm{keV})$  &  $1.30\pm0.01$ & -- & -- \\
$A_{{\mathrm{Mg}}} \ (\rm{ph \, s^{-1} \, cm^{-2}})$  &  $0.506^{+0.033}_{-0.031}$ & -- & -- \\
$\sigma_{{\mathrm{Mg}}} \ (\rm{keV})$  &  $0.424^{+0.004}_{-0.005}$ & -- & -- \\
$E_{\rm{Fe\,K\alpha_2}} \ (\rm{keV})$  &  $6.3970^{+0.0009}_{-0.0010}$ & $6.3959^{+0.0015}_{-0.0016}$ & $6.3905\pm{0.0008}$ \\
$A_{\rm{Fe\,K\alpha_2}} \ (\rm{ph \, s^{-1} \, cm^{-2}})$  &  $(1.54^{+0.01}_{-0.09})\times10^{-3}$ & $(1.14^{+0.20}_{-0.19})\times 10^{-3}$ & $(0.71\pm{0.10})\times10^{-3}$ \\
$\sigma_{\rm{Fe\,K\alpha_2}} \ (\rm{keV})$  & $(9.51^{+0.40}_{-0.52})\times10^{-3}$ & $(8.39^{+0.89}_{-0.96})\times 10^{-3}$ & $(4.80^{+0.78}_{-0.76})\times 10^{-3}$ \\
$E_{\rm{Fe\,K\alpha_1}} \ (\rm{keV})$  &  $6.4048\pm0.0007$  & $6.4050\pm0.0006$ & $6.4041\pm0.0004$
 \\
$A_{\rm{Fe\,K\alpha_1}} \ (\rm{ph \, s^{-1} \, cm^{-2}})$  &  $(0.36^{+0.10}_{-0.09})\times10^{-3}$ & $(0.48\pm0.15)\times 10^{-3}$ & $(1.14^{+0.09}_{-0.10})\times10^{-3}$ \\
$\sigma_{\rm{Fe\,K\alpha_1}} \ (\rm{keV})$  &  $(1.81^{+0.59}_{-0.55})\times10^{-3}$ & $(2.60^{+0.83}_{-0.71})\times 10^{-3}$ & $(4.12^{+0.38}_{-0.37})\times 10^{-3}$ \\
$E_{\rm{Fe\,K\beta}} \ (\rm{keV})$  & $7.0637^{+0.0027}_{-0.0026}$ & $7.0634^{+0.0010}_{-0.0011}$ & $7.0630\pm0.0014$ \\
$A_{\rm{Fe\,K\beta}} \ (\rm{ph \, s^{-1} \, cm^{-2}})$  & $(0.21\pm0.04)\times10^{-3}$ & $(0.09\pm 0.03)\times 10^{-3}$ & $(0.23^{+0.04}_{-0.03})\times10^{-3}$ \\
$\sigma_{\rm{Fe\,K\beta}} \ (\rm{keV})$  & $(7.10^{+2.51}_{-2.65})\times10^{-3}$ & $(0.84^{+1.54}_{-0.82})\times 10^{-3}$ & $(8.04^{+1.35}_{-1.48})\times10^{-3}$ \\
$E_{\rm{Ni\,K\alpha}} \ (\rm{keV})$  & $7.4513\pm0.0001$ & $7.4475^{+0.0034}_{-0.0035}$ & $7.4771\pm0.0019$ \\
$A_{\rm{Ni\,K\alpha}} \ (\rm{ph \, s^{-1} \, cm^{-2}})$  & $\leq 0.04\times10^{-3}$ & $\leq 3\times10^{-6}$ & $(0.06^{+0.03}_{-0.02})\times10^{-3}$ \\
$\sigma_{\rm{Ni\,K\alpha}} \ (\rm{keV})$  & $\leq 0.26\times10^{-3}$ & $\leq 0.63\times10^{-3}$ & $(3.63^{+2.85}_{-2.65})\times10^{-3}$ \\
C-stat$^{(a)}$  &  $1.37$ & $1.21$ & $1.17$ \\
\hline
\multicolumn{4}{p{0.7\linewidth}}{$^{(a)}$ The parameters derived from our variational inference analysis with \texttt{jaxspec} were loaded in ISIS to compute the equivalent C-stat to allow for frequentist goodness of fit assessment.} \\ 
\end{tabular}
\end{small}
\end{center}
\end{table}

\end{appendix}

\end{document}